\documentclass{ws-ijmpa}
\begin{document}

\markboth{V.~M.~Mostepanenko, V.~B.~Bezerra, G.~L.~Klimchitskaya
\& C.~Romero}{New Constraints on
Yukawa-Type Interactions from the Casimir Effect}

\catchline{}{}{}{}{}
%
%%%%%%%%%%%%%%%%%%%%%%%%%%%%%%%%%%%%%%%%%%%%%%%%%%%%%%%%%%%%%%%%%%%%

\title{NEW CONSTRAINTS ON YUKAWA-TYPE INTERACTIONS FROM
THE CASIMIR EFFECT}

\author{
V.~M.~MOSTEPANENKO
}

\address{Noncommercial Partnership
``Scientific Instruments'',
Tverskaya Street 11, Moscow,
103905, Russia {\protect \\}
Vladimir.Mostepanenko@itp.uni-leipzig.de
}

\author{V.~B.~BEZERRA}

\address{Department of Physics, Federal University of Para\'{\i}ba,
C.P.5008, CEP 58059--970, Jo\~{a}o Pessoa, Pb-Brazil}

\author{
G.~L.~KLIMCHITSKAYA
}

\address{North-West Technical University,
 Millionnaya Street 5, St.Petersburg,
191065, Russia
}

\author{C.~ROMERO}

\address{Department of Physics, Federal University of Para\'{\i}ba,
C.P.5008, CEP 58059--970, Jo\~{a}o Pessoa, Pb-Brazil}

\maketitle

\begin{history}
\received{16 November 2011}
\revised{Day Month Year}
\end{history}

\begin{abstract}
Measurements of the Casimir force are used to obtain
stronger constraints on the parameters of hypothetical
interactions predicted in different unification schemes
beyond the Standard Model.  We review new strong
constraints on the Yukawa-type interactions derived during the last two
years from recent experiments on
measuring the lateral Casimir force, Casimir force in
configurations with corrugated boundaries and the
Casimir-Polder force. Specifically, from measurements of the
lateral Casimir force compared with the exact theory the strengthening
of constraints up to a factor of 24 millions was achieved.
We also discuss further possibilities to strengthen constraints on the
Yukawa interactions from the Casimir effect.
\keywords{Yukawa-type interation; Casimir force; non-Newtonian gravity.}
\end{abstract}

\ccode{PACS numbers: 14.80.-j, 04.50.-h, 04.80.Cc, 12.20.Fv}

\section{Introduction}

It is common knowledge that Newton's gravitational law is not tested
experimentally with sufficient precision at separations below 0.1\,mm.
The smaller is the separation between the test bodies, the greater
is the correction to Newtonian gravity allowed by the experimental
data. For instance, at separation of $1\,\mu$m the existence of a
``correction'' that is nine orders of magnitude larger than the
magnitude of Newton's potential is not excluded experimentally.
So large addition scarcely can be called a correction.
Because of this it is conventional to speak about
non-Newtonian gravity.\cite{1}

The most offen discussed non-Newtonian gravity is of Yukawa-type.
In this case the complete gravitational potential, including both
the Newtonian and non-Newtonian contributions takes the form\cite{1}
\begin{equation}
V(r)=-\frac{Gm_1m_2}{r}\left(1+\alpha\,e^{-r/\lambda}\right),
\label{eq1}
\end{equation}
\noindent
where $m_1$ and $m_2$ are the masses of two pointlike bodies
separated by a distance $r$, $G$ is the Newtonian gravitational
constant and $\alpha,\,\lambda$ are the interaction constant and
interaction range of the additional force, respectively.

Modern physics provides two main reasons why the non-Newtonian
gravity of Yukawa-type should arise. The first of them is the exchange
of light elementary particles between separate atoms of two
macrobodies. Many particles, such as, scalar axion,\cite{2}
graviphoton,\cite{3} dilaton,\cite{4} goldstino,\cite{5} and
moduli,\cite{6} among others, are predicted in different extensions
of the Standard Model. The exchange of light particles of mass
$m=\hbar/(\lambda c)$ generates the effective Yukawa-type potential
shown by the second term on the right-hand side of
 Eq.~(\ref{eq1}). At the present time
light elementary particles are regarded as possible constituents
of dark matter and dark energy.

Another prediction of Yukawa-type additions to Newton's
gravitational law comes from extra-dimensional theories with
low-energy compactification scale.\cite{7}\cdash\cite{9}
According to these theories the compactification energy may be as
low as the extra-dimensional Planck energy
\begin{equation}
E_{\rm Pl}^{(D)}=\left(\frac{\hbar^{N+1}c^{N+5}}{G_D}
\right)^{\frac{1}{2+N}}\sim 1\,\mbox{TeV}.
\label{eq2}
\end{equation}
\noindent
Here $D=4+N$, $N$ is the number of extra dimensions compactified
at some small length $R_{\ast}$, $G_D=G\Omega_N$ is the gravitational
constant in the extended $D$-dimensional space-time and
$\Omega_N\sim R_{\ast}^N$.
When $N=0$ one obtains from (\ref{eq2}) the usual Planck energy
$E_{\rm Pl}=(\hbar c^5/G)^{1/2}\sim 10^{19}\,$GeV.
The characteristic size of the $N$-dimensional manifold is
given by\cite{9}
\begin{equation}
R_{\ast}\sim\frac{\hbar c}{E_{\rm Pl}^{(D)}}
\left[\frac{E_{\rm Pl}}{E_{\rm Pl}^{(D)}}
\right]^{\frac{2}{N}}\sim 10^{\frac{32-17N}{N}}\,\mbox{cm}.
\label{eq3}
\end{equation}

The presence of extra dimensions modifies the standard Newton
potential. Thus, at separations $r\gg R_{\ast}$ the resulting
gravitational potential takes the form (\ref{eq1}).
It is pertinent to compare the theoretical prediction (\ref{eq3})
with available experimental data. Thus, for $N=1$
Eq.~(\ref{eq3}) leads to $R_{\ast}\sim 10^{15}\,$cm.
Such  a large extra dimension is excluded by solar system tests
of Newtonian gravitational law.\cite{1} For $N=2$ and 3
the sizes of extra dimensions are $R_{\ast}\sim 1\,$mm and
$\sim 5\,$nm, respectively. Keeping in mind that
$\lambda\sim R_{\ast}$, these predictions fall within the range where
constraints on non-Newtonian gravity are not so stringent.

Constraints on the Yukawa-type corrections to Newton's law were
traditionally obtained from the E\"{o}tvos- and Cavendish-type
experiments. For this purpose the Yukawa interaction energy and
force between two test bodies separated by a distance $a$ were
obtained by integration of the second term in Eq.~(\ref{eq1})
over their volumes and subsequent negative differentiation with
respect to $a$:
\begin{equation}
E_{\rm Yu}(a)=-G\alpha \rho_1\rho_2
\int_{V_1}\int_{V_2}
d\mbox{\boldmath$r$}_1d\mbox{\boldmath$r$}_2
\frac{e^{-|\mbox{\boldmath$r$}_1-
\mbox{\boldmath$r$}_2|/\lambda}}{|\mbox{\boldmath$r$}_1-
\mbox{\boldmath$r$}_2|}, \quad
F_{\rm\, Yu}(a)=-\frac{\partial E_{\rm Yu}(a)}{\partial a},
\label{eq4}
\end{equation}
\noindent
where $\rho_{1},\,\rho_{2}$ are the matter
densities. The obtained constraints are discussed in detail in the
literature.\cite{1,10,11}
The strength of these constraints decreases with decreasing $\lambda$.
Thus, for the range of the smallest $\lambda$ such that
$4.7\,\mu\mbox{m}<\lambda<9\,\mu$m the maximum allowed value of $\alpha$
varies\cite{12,12a}
between $3.2\times 10^7$ and $10^5$. For $\lambda<4.7\,\mu$m the strongest
constraints on the Yukawa-type interactions follow from measurements
of the Casimir force\cite{11,13} which becomes the dominant background
force at short separations between the test bodies. Note that the
possibility to constrain the non-Newtonian gravity from measurements
of the van der Waals and Casimir forces was proposed long ago for
the Yukawa\cite{14} and power-type\cite{15} corrections to Newton's law.
A review of constraints on the Yukawa-type interactions from the Casimir
effect obtained up to 2009 is available.\cite{11}

In this paper we review new constraints on non-Newtonian gravity obtained
from measurements of the Casimir force during the last two years.
During this period a major progress was achieved. Specifically,
at the shortest separation range, the strength of high confidence constraints
was increased up to a factor of $2.4\times 10^7$.
We also consider the prospects of further strengthening these constraints
using new experiments on the Casimir force. In Sec.~2 we present constraints
obtained from measuring the normal Casimir force between test bodies with
smooth surfaces. Section 3 is devoted to constraints following from
measurements of the normal Casimir force between test bodies with corrugated
surfaces. The constraints following from measurements of the lateral
Casimir force are considered in Sec.~4. The constraints obtained from
measurements of the thermal Casimir-Polder force are discussed in Sec.~5.
Section 6 is devoted to possibilities to further strengthening constraints
on the Yukawa interaction using new experimental configurations.
Our conclusions are contained in Sec.~7.

\section{Constraints from the Normal Casimir Force Between Test Bodies
with Smooth Surfaces}

Here we remind the methodology on how constraints on the Yukawa interactions
are obtained from measurements of the normal Casimir force (i.e., acting
in the normal direction to the surface). We also present the constraints
obtained up to 2009 to be compared with stronger and (or) more
recent constraints. As the first such constraint we discuss the one
obtained\cite{15a} from new torsion pendulum experiment.\cite{15b}

The measured quantities are the Casimir force $F_C(a,T)$ in a sphere-plate
geometry (experiment using  an atomic force microscope\cite{16})
and its gradient $F_C^{\prime}(a,T)=\partial F_C(a,T)/\partial a$
(experiment using a micromachined oscillator\cite{17,18}).
The high-confidence constraints determined at a 95\% confidence level
are obtained from the inequalities
\begin{equation}
\left|F_{\rm Yu}(a)\right|\leq \Xi_{F_C}(a),\qquad
\left|F_{\rm Yu}^{\prime}(a)\right|\leq \Xi_{F_C^{\prime}}(a).
\label{eq5}
\end{equation}
\noindent
Here $F_{\rm Yu}(a)$ is the Yukawa force between a sphere (or
a spherical lens)
and a plate calculated using Eq.~(\ref{eq4}).
The quantity $\Xi_{F_C}$ ($\Xi_{F_C^{\prime}}$) is the half-width of the
confidence interval for the difference between theoretical and
measured Casimir force (force gradient) determined at a 95\% confidence
level. The confidence intervals should not be confused with the force
residuals, i.e., the differences between the computed and
measured forces (although at least 95\% of force residuals belong to the
confidence intervals).

It should be emphasized that $\Xi_{F_C}$ and $\Xi_{F_C^{\prime}}$
have the meaning of absolute errors. Because of this, they can be
determined with only two or three significant figures independently of
their values, i.e., with the relative error of about 0.5\%.
{}From this it follows that the Yukawa force $F_{\rm Yu}(a)$ needs not
 be calculated with a higher precision (even if the
measurement of the Casimir force was more precise).
It was shown\cite{19,20} that the Yukawa force in sphere-plate geometry
can be calculated both precisely by Eq.~(\ref{eq4}) and using the
simplified formulation of the proximity force approximation
\begin{equation}
F_{\rm Yu}^{sp}(a)=2\pi RE_{\rm Yu}^{pp}(a),
\label{eq6}
\end{equation}
\noindent
where $R$ is the sphere radius and $E_{\rm Yu}^{pp}(a)$ is the Yukawa energy
per unit area in the configuration of two parallel plates.
With the experimental parameters used\cite{17,18}
the results of both calculations coincide.\cite{20}

It was claimed\cite{21} that one and the same experiment cannot be used
simultaneously to exclude or confirm some theory of the Casimir force
and to place constraints on the Yukawa force from the measure of
agreement between experiment and theory. This opinion would be warranted
only if the difference between the excluded and confirmed theories
of the Casimir force can be modeled by the Yukawa force.
This is, however, not so in all experiments discussed here.
Therefore, it became a common practice in the literature to compare
the measured data with different theories and, after a selection
of the theory in agreement with the data is made, to obtain constraints
on the Yukawa interaction.

In Fig.~1 we present constraints on the parameters of Yukawa-type
interaction $(\lambda,\,\alpha)$ following\cite{18,22} from
measurements of the normal Casimir force between Au surfaces by means
of an AFM\cite{16} (line 1), from measuring the gradient of the Casimir
force between similar surfaces (or, equivalently, the Casimir pressure
between two Au-coated parallel plates) by means of a micromachined
oscillator\cite{17,18} (line 2), and from the Casimir-less
experiment,\cite{23}\cdash\cite{23b} where the contribution of
the Casimir force was made equal to zero (line 3). The constraints for
larger $\lambda$ were obtained\cite{24} from the torsion-pendulum
experiment\cite{25} of 1997 (line 4). The constraints from
the torsion-pendulum experiment\cite{26} of 2009
are shown by the line 5. The dashed
line 6 in Fig.~1 shows the constraints following from the gravitational
experiment of Cavendish-type\cite{12,12a} which leads to the
strongest constraints in this interaction range. At larger $\lambda$ all
the strongest constraints follow from the gravitational experiments as
discussed in Sec.~1. In this and in all other figures below, the
allowed values of $\lambda$ and $\alpha$ in the $(\lambda,\,\alpha)$-plane
lie beneath the respective lines.

%%%%%%%%___FIGURE_1___%%%%%%%
\begin{figure*}[t]
\vspace*{-5.5cm}
\centerline{\hspace*{4.cm}\psfig{file=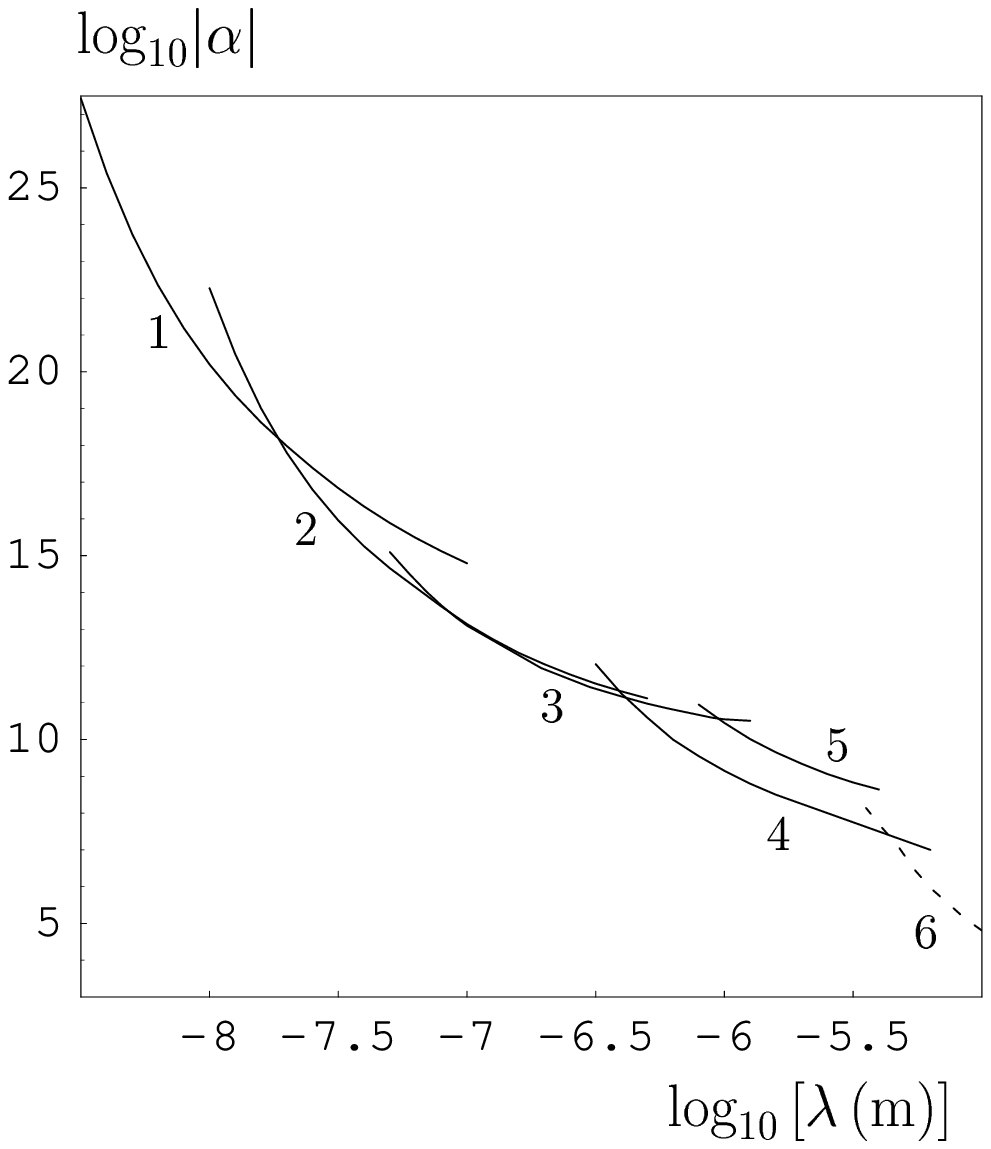,width=10cm}}
\vspace*{-3.2cm}
\caption{The constraints on Yukawa-type interactions
obtained from measurements of the normal Casimir force
 between a sphere (spherical lens) and a plate
by means of an AFM (line 1),   or the
Casimir pressure using a micromachined
oscillator (line 2), from the Casimir-less experiment (line 3),
from the torsion-pendulum experiments of 1997 (line 4) and
2009 (line 5). The dashed line 6 shows the strongest constraints
following from gravitational experiments (see text for further discussion).
}
\end{figure*}
%%%%%%%%%%%%%%%

Now let us begin the consideration of new information concerning
the constraints shown in Fig.~1 obtained in the last two years.
We start with the constraints\cite{15a} obtained from the
measurement data of recent torsion-pendulum experiment,\cite{15b}
which is similar to previous experiments using the torsion
pendulum,\cite{25,26} but arrives at quite different conclusions.
This experiment was performed in the geometry of a spherical lens
of $R=15.6\,$cm radius of curvature near a plate, both coated with
Au. The total force of nonestablished origin (presumably caused by
large electrostatic patches) between the lens and the plate was up
to a factor of 10 larger than the Casimir force even when the
residual electric force was compensated. This total force was measured
over a wide range of separations from 0.7 to $7.3\,\mu$m.
The Casimir force was extracted using the fitting procedure
with two fitting parameters.
It was found in agreement with the Drude model and in contradiction with
the plasma model (contrary to the results of the previous
torsion-pendulum experiments\cite{25,26} and experiments using a micromachined
oscillator\cite{17,18}). The constraints obtained\cite{15a} are shown
in Fig.~2 by the line 4a. The lines 3, 4, and 6 are the same, as explained
in the caption of Fig.~1 and respective discussion in the text.
Specifically, line 4 shows the constraints obtained\cite{24} from the
first torsion-pendulum experiment.\cite{25}

%%%%%%%%___FIGURE_2___%%%%%%%
\begin{figure*}[t]
\vspace*{-8.5cm}
\centerline{\hspace*{4.cm}\psfig{file=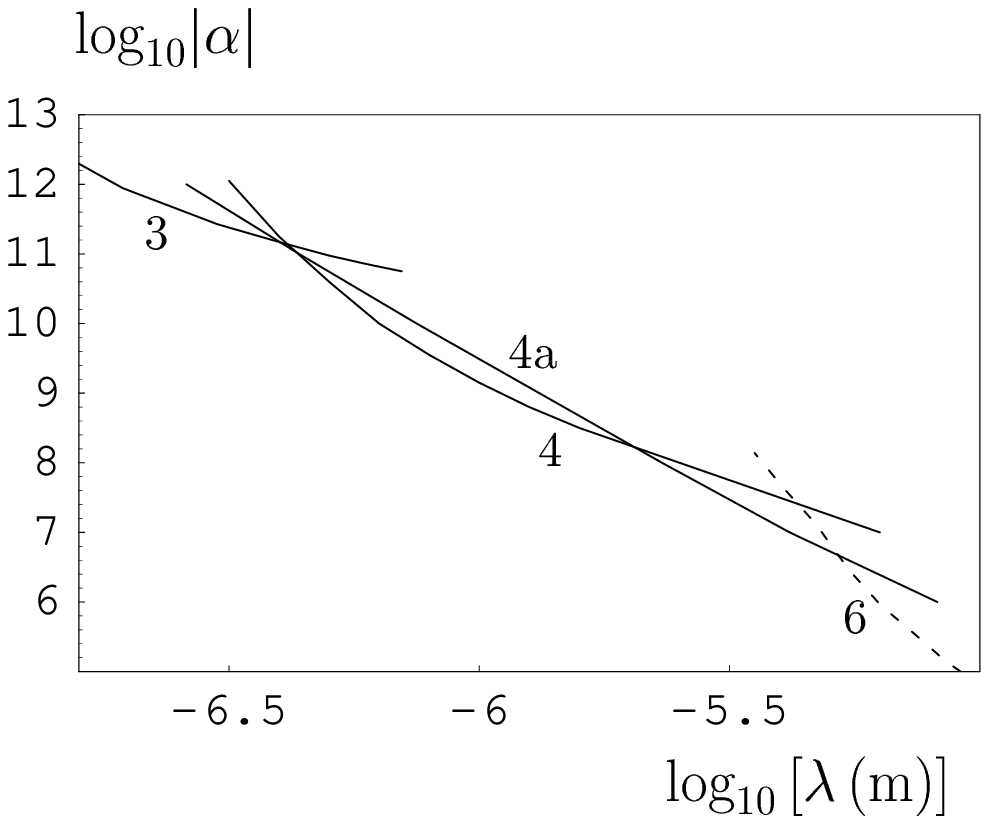,width=12cm}}
\vspace*{-3.8cm}
\caption{The constraints on Yukawa-type interactions
obtained from the torsion-pendulum experiments of 2011 (line 4a) and
1997 (line 4), and from the Casimir-less experiment (line 3).
The dashed line 6 shows the strongest constraints
following from gravitational experiments.
}
\end{figure*}
%%%%%%%%%%%%%%%

The major problem of the recent torsion-pendulum experiment\cite{15b}
is that at separations above $3\,\mu$m the Casimir force obtained
after the subtraction agrees much better not with the Drude
model, as claimed, but with
the plasma model.\cite{27} As to separations below $3\,\mu$m,
the results of Ref.~\refcite{15b} were shown to be not reliable
because of imperfections invariably present on surfaces of large
lenses which were not taken into account.\cite{28}
As can be seen in Fig.~2, the constraints shown by the lines 4 and 4a
are almost of the same strength in spite of the fact that they follow
from the two experiments in mutual contradiction.
In any case it should be concluded that all constraints following from
the torsion-pendulum experiments using large glass lenses (lines 4 and 5
in Fig.~1 and lines 4 and 4a in Fig.~2) are lacking reliability due to
the problems indicated in Ref.~\refcite{28}. They are not as reliable
as the constraints shown by the lines 1--3 which are deternimed at a 95\%
confidence level.

\section{Constraints from the Normal Casimir Force Between Test Bodies
with Corrugated Surfaces}

These constraints on the Yukawa-type interaction were obtained\cite{29}
from the results of experiment\cite{30} measuring the gradient of
the normal Casimir force between an Au-coated sphere above a Si plate
covered with corrugations of trapezoidal shape (see also Ref.~\refcite{30a}).
The experimental
configuration is schematically shown in Fig.~3(a).
%%%%%%%%___FIGURE_3___%%%%%%%
\begin{figure*}[t]
\vspace*{-6.7cm}
\centerline{\hspace*{4.5cm}\psfig{file=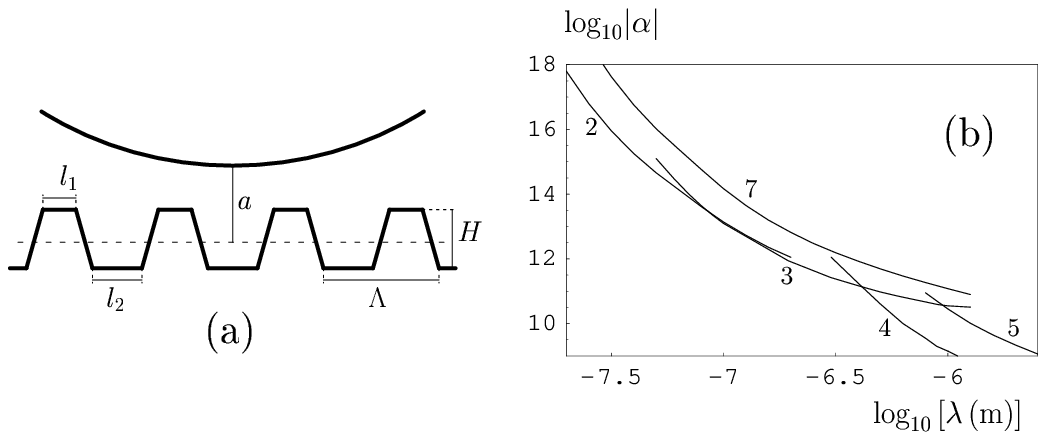,width=23cm}}
\vspace*{-21.2cm}
\caption{(a) The configuration of the experiment on measuring
the normal Casimir force between a smooth sphere and a plate
covered with trapezoidal corrugations.
(b) The constraints on Yukawa-type interactions obtained from the
experiment using a trapezoidally corrugated plate (line 7),
 a micromachined oscillator (line 2), from the Casimir-less
experiment (line 3) and from the torsion-pendulum experiments
of 1997 (line 4) and 2009 (line 5).
}
\end{figure*}
%%%%%%%%%%%%%%%
The gradient of Yukawa force in this configuration was found
in the form\cite{29}
\begin{equation}
F_{\rm Yu,corr}^{\prime}(a)=-\left(\frac{p_1}{\lambda}+
\frac{p_3}{H}\right)F_{\rm Yu}^{sp}(a-H_1)-
\left(\frac{p_2}{\lambda}-
\frac{p_3}{H}\right)F_{\rm Yu}^{sp}(a+H_2).
\label{eq7}
\end{equation}
\noindent
Here $p_i=l_i/\Lambda$ ($i=1,\,2$), $p_3=1-p_1-p_2$, and
\begin{equation}
H_i=\frac{\Lambda-l_i}{2\Lambda-l_1-l_2}.
\label{eq8}
\end{equation}
\noindent
The Yukawa force between an Au-coated sphere of radius $R$ and a smooth
plate is given by\cite{20}
\begin{equation}
F_{\rm Yu}^{sp}(z)=-4\pi^2R\alpha G\lambda^3e^{-z/\lambda}\rho_p
\left[\rho_{\rm Au}\Phi_1\left(\frac{\lambda}{R}\right)-
(\rho_{\rm Au}-\rho_s)
\Phi_2\left(\frac{\lambda}{R},\frac{\Delta_{\rm Au}}{R}\right)
e^{-\Delta_{\rm Au}/\lambda}\right],
\label{eq9}
\end{equation}
\noindent
where $\rho_p,\>\rho_s,\>\rho_{\rm Au}$ are the densities of the plate,
sphere, and Au, respectively, $\Delta_{\rm Au}$ is the thickness
of the Au coating on the sphere and the functions $\Phi_1(x)$ and
$\Phi_2(x,y)$ are defined as
\begin{eqnarray}
&&
\Phi_1(x)=1-x+(1+x)e^{-2/x},
\nonumber \\
&&
\Phi_2(x,y)=1-x-y+(1+x-y)e^{-2(1-y)/x}.
\label{eq10}
\end{eqnarray}
\noindent
Note that under the conditions
$x=\lambda/R\ll 1$ and $y=\Delta_{\rm Au}/R\ll 1$ we have
$\Phi_{1,2}\to 1$ and Eq.~(\ref{eq9}) takes the form
\begin{equation}
F_{\rm Yu,PFA}^{sp}(a)=2\pi RE_{\rm Yu}(a),
\label{eq11}
\end{equation}
\noindent
where the Yukawa energy per unit area of the two parallel plates
is given by the following expression:
\begin{equation}
E_{\rm Yu}(a)=-2\pi\alpha G\lambda^3e^{-a/\lambda}\rho_p
\left[\rho_{\rm Au}-(\rho_{\rm Au}-\rho_s)e^{-\Delta_{\rm Au}/\lambda}
\right].
\label{eq12}
\end{equation}
\noindent
Equations (\ref{eq11}) and (\ref{eq12}) were used in calculations of the
Yukawa force in the experiments considered in Sec.~2. They lead to the
same results for the constraints as obtained from a more exact
Eq.~(\ref{eq9}).

In Fig.~3(b) the constraints on the parameters of Yukawa force following
from the experimental data\cite{30} are shown by the line 7.
For comparison purposes, in the same figure the lines 2--5 present the
same constraints as in Figs.~1 and 2.
As can be seen in Fig.~3(b), the constraints shown by the line 7 are
slightly weaker than the constraints shown by lines 2 and 3.
At $\lambda=1.26\,\mu$m, where the minimum difference between the
constraints of lines 2 and 3, on the one hand, and line 7, on the other
hand, is achieved, the constraints of line 7 are weaker by a factor of 2.4.
The reason is that the density of Si is smaller that the density of Au
deposited on both test bodies in the respective
experiments.\cite{17,18,23}\cdash\cite{23b}
It is pertinent to note, however, that line 7 in Fig.~3(b) is in  a very
good qualitative agreement with lines 2 and 3 and, thus, provides
confirmation to previously obtained constraints.
In fact this experiment alone covers the interaction ranges of two
previously performed
experiments. If the material of the corrugated plate (Si) were replaced
by Au, stronger by a factor 8.3 constraints than those shown by line 7
would be obtained (see Sec.~7 where the possibilities to obtain stronger
constraints are discussed).

\section{Constraints from the Lateral Casimir Force}

The lateral Casimir force acting between the sinusoidally corrugated
surfaces of a sphere and a plate has long been demonstrated
and the measurement data were compared
with the theory using the PFA.\cite{31,32}
Recently more precise measurements of the lateral force were
performed and compared with the exact scattering theory.\cite{33,34}
The measure of agreement between the data and theoretical predictions
was used\cite{35} to obtain stronger constraints on the Yukawa-type
interaction. For this purpose the lateral Yukawa force was calculated
in the configuration of a sinusoidally corrugated sphere above a
sinusoidally corrugated plate, both coated with Au. The period $\Lambda$
of corrugations on both surfaces was common, but the corrugation
amplitudes $A_1$ and $A_2$ were different.
The cause of the appearance of the lateral Casimir force is the phase shift
$\varphi$ between corrugations on the sphere and the plate.

The obtained expression for the lateral Casimir force is the
following:\cite{35}
\begin{equation}
F_{\rm Yu, lat}^{sp,\rm corr}(a,\varphi)=8\pi^3G\alpha\lambda^3
\Psi(\lambda)e^{-a/\lambda}\frac{A_1A_2}{b\Lambda}
I_1\left(\frac{b}{\lambda}\right)\sin\varphi.
\label{eq13}
\end{equation}
\noindent
Here, $I_n(z)$ is the Bessel function of imaginary argument and
the quantity $b$ is given by
\begin{equation}
b\equiv b(\varphi)=(A_1^2+A_2^2-2A_1A_2\cos\varphi)^{1/2}.
\label{eq14}
\end{equation}

The function $\Psi(\lambda)$ in (\ref{eq13}) is defined with account
of the layer structure of the sphere and the plate. The corrugated plate
of density $\rho_p$ was covered with an Au layer of thickness
$\Delta_{{\rm Au},p}$. As to the sphere with density $\rho_s$,
it was first coated with a Cr layer of density $\rho_{\rm Cr}$ and
thickness $\Delta_{\rm Cr}$, and then with an Au layer with density
$\rho_{\rm Au}$ and thickness
$\Delta_{{\rm Au},s}$.
Taking all this into account, the function $\Psi(\lambda)$ can be presented
in the form\cite{35}
\begin{eqnarray}
\Psi(\lambda)&=&\left[\rho_{\rm Au}-(\rho_{\rm Au}-\rho_p)
e^{-\Delta_{{\rm Au},p}/\lambda}\right]\left[\rho_{\rm Au}\Phi(R,\lambda)
-(\rho_{\rm Au}-\rho_{\rm Cr})\Phi(R-\Delta_{{\rm Au},s},\lambda)
\vphantom{e^{-\Delta_{{\rm Au},s}/\lambda}}
\right.
\nonumber \\
&&\left.
\times e^{-\Delta_{{\rm Au},s}/\lambda}
-(\rho_{\rm Cr}-\rho_{s})\Phi(R-\Delta_{{\rm Au},s}-\Delta_{\rm Cr},\lambda)
e^{-(\Delta_{{\rm Au},s}+\Delta_{\rm Cr})/\lambda}\right],
\label{eq15}
\end{eqnarray}
where the function $\Phi(x,\lambda)$ is defined as
\begin{equation}
\Phi(x,\lambda)=x-\lambda+(x+\lambda)e^{-2x/\lambda}.
\label{eq16}
\end{equation}

%%%%%%%%___FIGURE_4___%%%%%%%
\begin{figure*}[t]
\vspace*{-5.6cm}
\centerline{\hspace*{4.cm}\psfig{file=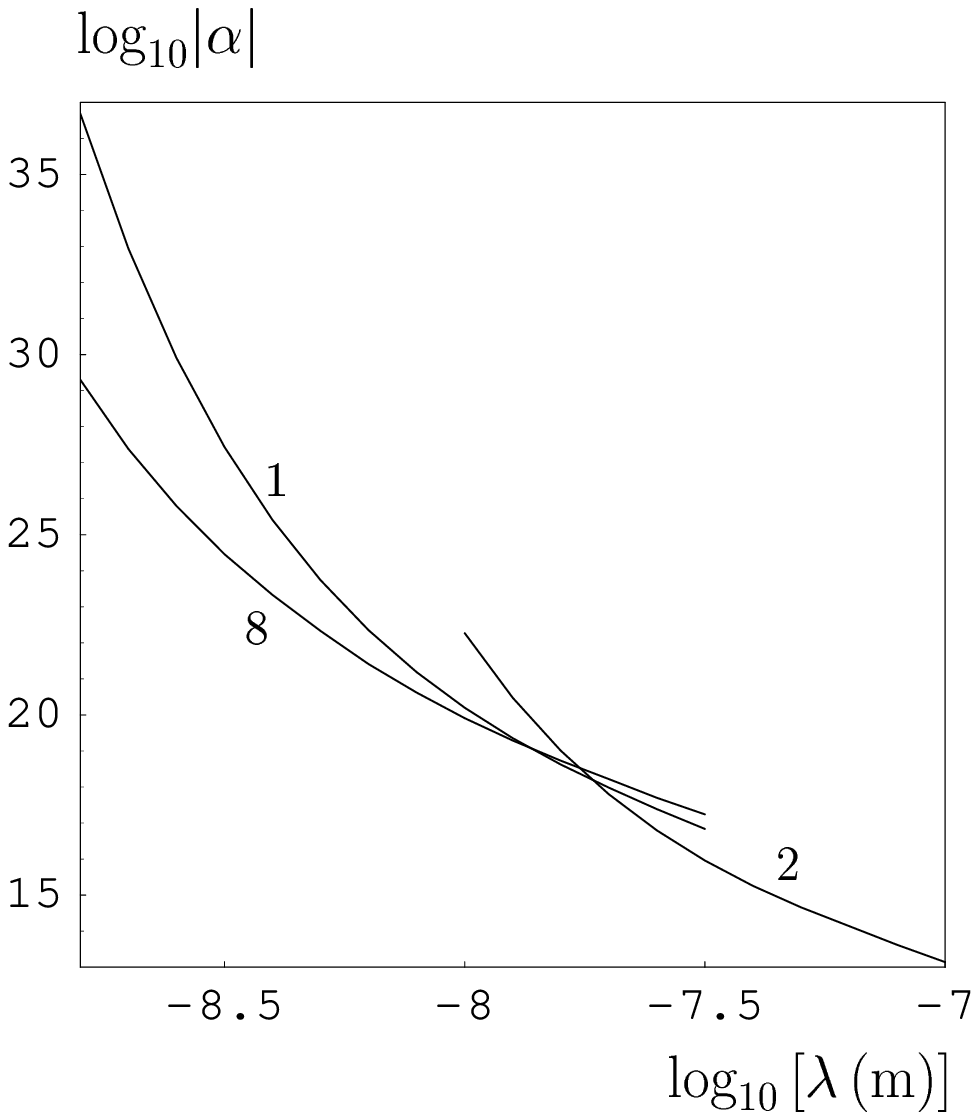,width=10cm}}
\vspace*{-3.2cm}
\caption{The constraints on Yukawa-type interactions
obtained from measurements of the lateral Casimir
force between corrugated surfaces of a sphere and a plate (line 8),
from measurements of the normal Casimir force
by means of an AFM (line 1) and of the Casimir pressure by means of
 a micromachined oscillator (line 2).
}
\end{figure*}
%%%%%%%%%%%%%%%

The constraints on the parameters of the Yukawa interaction obtained
using Eqs.~(\ref{eq13})--(\ref{eq16}) from the measure of agreement
between experiment and theory for the lateral Casimir force are
shown in Fig.~4 by line 8. Note that these constraints
are determined at the same 95\% confidence level as the total
experimental errors in measurements of the lateral Casimir
force.\cite{33,34} The lines 1 and 2 in Fig.~4 are the same as in
previous figures. They indicate the strongest high confidence
constraints obtained so far. As is seen in Fig.~4, the constraints
indicated by line 8 are the strongest over the interaction range
from 1.6 to 14\,nm. Remarkably, the largest strengthening of
previously known constraints shown by line 1 achieves a factor of
$2.4\times 10^7$ at $\lambda=1.6\,$nm.
The physical reason for such strong strengthening of the
constraints from the experiment with corrugated surfaces is that at
a separation, for instance, of $a=121\,$nm between the mean levels of
corrugations, the distance between the closest points of the
surfaces becomes as small as 22\,nm.
Note that for $\lambda$ below 1\,nm the strongest constraints on the
Yukawa interaction are obtained from precision atomic physics.\cite{36}

\section{Constraints from the Thermal Casimir-Polder Force}

The thermal Casimir-Polder force between ${}^{87}$Rb atoms and SiO${}_2$
plate in the separation region from 6.8 to $11\,\mu$m was
determined\cite{37} from the fractional frequency shift $\gamma_z(a)$
of Bose-Einstein condensate oscillations in the direction perpendicular
to the plate. The condensate confined in a magnetic trap was
characterized by the proper frequency $\omega_{0z}$ and by the
Thomas-Fermi radius $R_z$. The experimental data for $\gamma_z$ were
found\cite{37} to be in agreement with theory disregarding
conductivity of SiO${}_2$ plate at a constant current (dc conductivity).
It was shown\cite{38} also that the data
exclude the theory taking into account dc
conductivity of SiO${}_2$.
The constraints on the Yukawa interaction were obtained\cite{35}
from the measure of agreement with theory disregarding   dc
conductivity of SiO${}_2$. For this purpose the fractional frequency
shift due to the Yukawa interaction was found in the form\cite{35}
\begin{equation}
\gamma_{z,\rm Yu}(a)=\frac{15\pi G\lambda\rho_p}{8\omega_{0z}^2A_z}
\alpha e^{-a/\lambda}\Theta\left(\frac{R_z}{\lambda}\right)
I_1\left(\frac{A_z}{\lambda}\right),
\label{eq17}
\end{equation}
\noindent
where
\begin{equation}
\Theta(t)=\frac{16}{t^5}(t^2\sinh t-3t\cosh t+3\sinh t).
\label{eq18}
\end{equation}

%%%%%%%%___FIGURE_5___%%%%%%%
\begin{figure*}[t]
\vspace*{-5.7cm}
\centerline{\hspace*{4.cm}\psfig{file=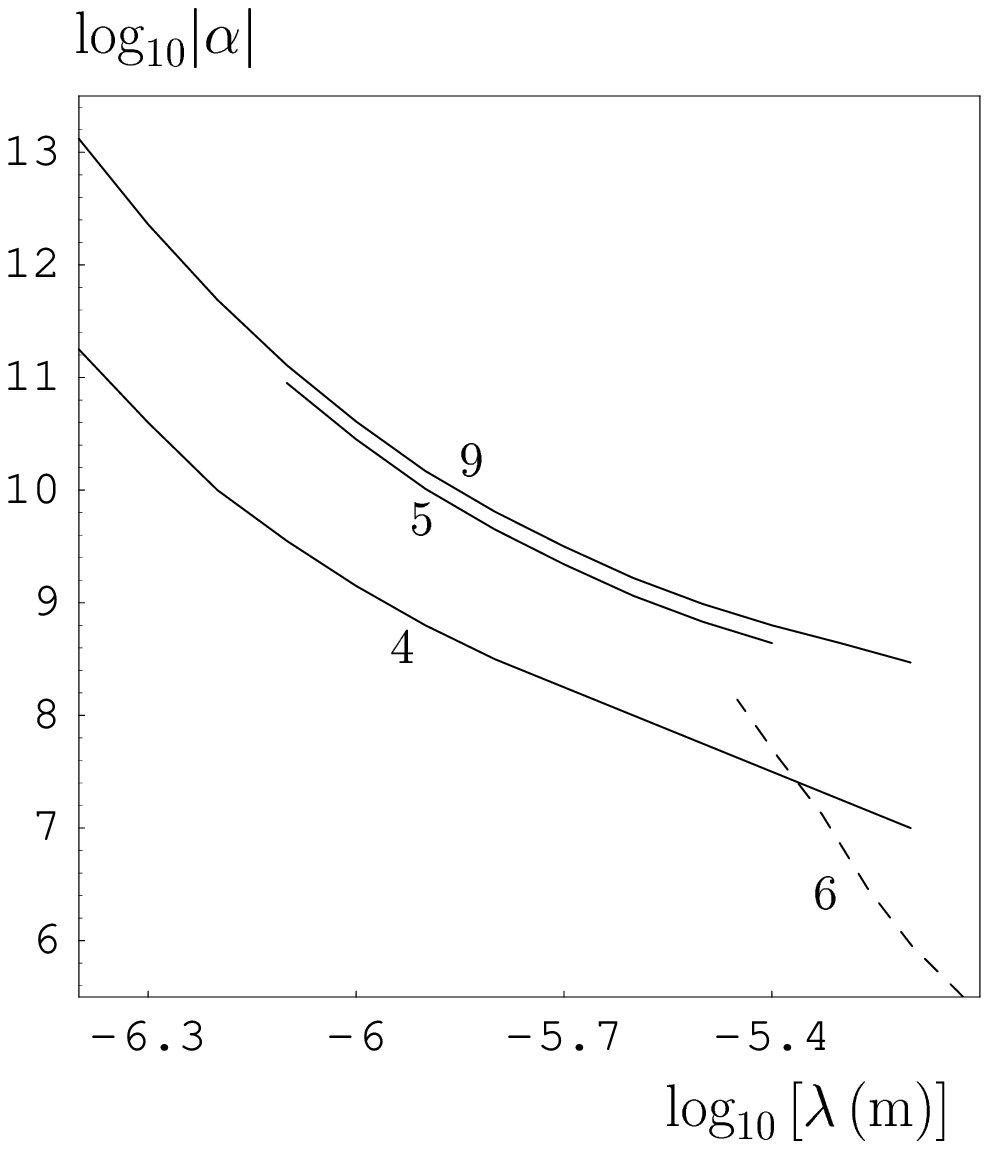,width=10cm}}
\vspace*{-3.2cm}
\caption{The constraints on Yukawa-type interactions
obtained from measurements of the Casimir-Polder
force between ${}^{87}$Rb atoms and a plate (line 9),
and from the torsion-pendulum experiments of 1997 (line 4)
and 2009 (line 5).
The dashed line 6 shows the strongest constraints
following from gravitational experiments.
}
\end{figure*}
%%%%%%%%%%%%%%%
The obtained constraints are shown by line 9 in Fig.~5.
They are characterized by the same 67\% confidence level, as the
experimental error in this experiment. The lines 4 and 5 present the
constraints obtained from torsion pendulum experiments.\cite{25,26}
As discussed in Sec.~2, there are reasons to believe that they are
not enough reliable. The line 6 shows the strongest constraints
following from gravitational experiments in this interaction range.
As can be seen in Fig.~5, the constraints of line 4 are from about 1
to 2 orders of magnitude stronger than the constraints of line 9.
The constraints of line 5 are slightly stronger than the constraints
of line 9 obtained from the measurement of Casimir-Polder force.
Nevertheless the advantage of the constraints shown by line 9, as
compared with constraints of lines 4 and 5, is that they are
obtained on the basis of an independent measurement and do not use
fitting parameters in the comparison between the measured data and theoretical
prediction.

\section{Possibilities to  Strengthen Constraints on the Yukawa
Interactions from Measurements of the Casimir Force}

In the foregoing we have considered new limits on the Yukawa-type
interaction obtained during the last two years from experiments on
measuring the Casimir force. The major strengthening up to a factor of
24 millions was achieved from measurement of the lateral Casimir force
between sinusoidally corrugated surfaces of a sphere and a plate.
Here we discuss further possibilities to strengthen constraints on the
Yukawa interaction using corrugated surfaces. We also discuss the
strengthening of constraints on the Yukawa interaction that can be
achieved from measuring the gradient of the Casimir force between a
microfabricated cylinder and a plate.

\subsection{Constraints from proposed measurements of the gradient
of the Casimir force between a smooth sphere and a plate with
trapezoidal corrugations}

First we consider the experiment using a smooth sphere and a plate with
trapezoidal corrugations considered in Sec.~4. The constraints on
Yukawa interaction obtained from this experiment are shown by line 7
in Fig.~3(b). As was discussed in Sec.~4, the reason why these
constraints are weaker than the previously obtained (lines 2 and 3) is that
the density of Si is smaller than the density of Au used as the plate
material in the experiments of Refs.~\refcite{17,18} and \refcite{23}.
Here we preserve the same experimental parameters, as were used in the
original experiment,\cite{30} but replace the Si corrugated plate with
a similar plate made of Au. Repeating calculations of the gradient
of Yukawa force by Eq.~(\ref{eq7}) but with $\rho_p=\rho_{\rm Au}$
we obtain\cite{29} the constraints shown by the dashed line in Fig.~6(a).
For comparison purposes, in Fig.~6(a) we also plot the same lines
2, 3, 4, 5, and 7 as in Fig.~3(b). As was mentioned in Sec.~4, the
constraints from experiment with an Au plate
covered with trapezoidal
corrugations are stronger by a factor $\rho_{\rm Au}/\rho_p\approx 8.3$
than the constraints of line 7. Within some interaction ranges, the
dashed line in Fig.~6(a) also demonstrates stronger constraints than
those shown by lines 3, 4, and 5. Thus, at $\lambda=0.94\,\mu$m,
the constraint shown by the dashed line is stronger by a factor of
3.8 than the constraints shown by lines 3 and 5.
At $\lambda=1.04\,\mu$m,
the dashed line gives by a factor of 3.5 stronger constraint
than that shown by lines 3 and 5.

\subsection{Constraints from proposed measurements of the normal
Casimir force between a smooth sphere and a sinusoidally
corrugated plate}

As was discussed in Sec.~2, line 1 in Fig.~1 shows constraints on the Yukawa
interaction following from measurements\cite{16} of the normal Casimir
force between smooth Au surfaces of a sphere and a plate using an AFM.
The constraints of line 1 were significantly improved from measurements
of the lateral Casimir force (see line 8 in Fig.~4).
The question arises whether the use of corrugated plate in an experiment
on measuring the normal Casimir force could lead to stronger constraints.
To answer this question, we have calculated the Yukawa force acting
between a smooth Au-coated sphere and a sinusoidally corrugated
Au plate\cite{35}
\begin{equation}
F_{\rm Yu}^{sp,\rm corr}(a)=-4\pi^2G\alpha\lambda^3\Psi(\lambda)
e^{-a/\lambda}I_0(A_1/\lambda),
\label{eq19}
\end{equation}
\noindent
where $A_1$ is the amplitude of corrugations on the plate and the
function $\Psi(\lambda)$ is defined in Eq.~(\ref{eq15}).
Note that the sphere is assumed to be coated with only one layer
of Au. This means that one should put $\Delta_{\rm Cr}=0$ in (\ref{eq15}).
The quantity $\Xi_{F_C}(a)$ from Eq.~(\ref{eq5}) was calculated for
this experiment in Ref.~\refcite{39}. In the computations of constraints
the experimental parameters of the two experiments\cite{16,33,34}
were used.

%%%%%%%%___FIGURE_6___%%%%%%%
\begin{figure*}[t]
\vspace*{-7.cm}
\centerline{\hspace*{4.5cm}\psfig{file=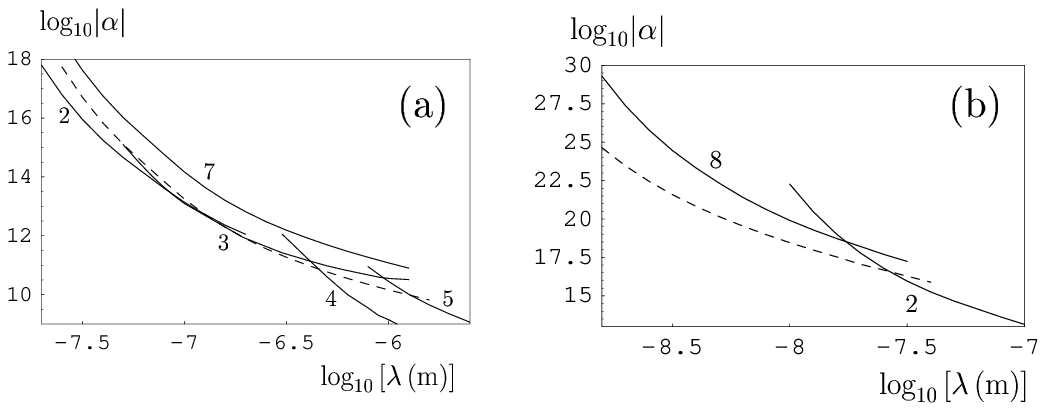,width=23cm}}
\vspace*{-21.2cm}
\caption{(a) The prospective constraints on Yukawa-type interactions
 from measurements of the gradient of the Casimir
force between an Au-coated sphere and Au plate covered with
trapezoidal corrugations are shown by
the dashed line. Lines 2, 3, 4, 5, and 7
are the same as in Fig.~3(b).
(b) The prospective constraints
 from measurements of the normal Casimir
force between an Au-coated sphere and Au plate covered with
sinusoidal corrugations  are shown by the dashed line. Lines 2 and 8
are the same as in Fig.~4.
}
\end{figure*}
%%%%%%%%%%%%%%%

The prospective constraints that can be obtained from this experiment are
shown by the dashed line in Fig.~6(b). Lines 8 and 2 show constraints
obtained from measurements of the lateral Casimir force using an AFM and
the Casimir pressure using a micromechanical oscillator, respectively.
As can be seen in Fig.~6(b), the dashed line provides a significant
strengthening of the best constraints shown by line 8.
Specifically, the largest strengthening by a factor of $4.5\times 10^4$
occurs at $\lambda=1.6\,$nm. Thus, the use of corrugated surfaces is
very prospective for constraining the Yukawa-type interactions.

\subsection{Prospects of using the configuration of a microfabricated
cylinder above a plate}

Recently it was proposed to measure the gradient of the Casimir force
by means of a micromechanical oscillator in the configurations of a
circular\cite{40} or elliptic\cite{41} microfabricated cylinder and a plate.
These configurations present some advantages in comparison with
configurations of two parallel plates and a sphere above a plate.
The Yukawa force acting between a plate and a microfabricated cylinder
of length $L$ and radius $R$ can be presented in the form\cite{42}
\begin{equation}
F_{\rm Yu}^{cp}(a)=-4\pi^2G\rho_1\rho_2\alpha\lambda^2LRe^{-(R+a)/\lambda}
I_1\left(\frac{R}{\lambda}\right),
\label{eq20}
\end{equation}
\noindent
where $\rho_1$ and $\rho_2$ are the densities of the cylinder and plate
materials and it is assumed that the thickness of the plate is much
larger than $\lambda$. {}From Eq.~(\ref{eq20}) for the gradient of the
Yukawa force one obtains
\begin{equation}
\frac{\partial F_{\rm Yu}^{cp}(a)}{\partial a}
=4\pi^2G\rho_1\rho_2\alpha\lambda LRe^{-(R+a)/\lambda}
I_1\left(\frac{R}{\lambda}\right).
\label{eq21}
\end{equation}
\noindent
This result was generalized\cite{42} for the case when there are two
thin layers of different densities covering both the cylinder and
the plate.

%%%%%%%%___FIGURE_7___%%%%%%%
\begin{figure*}[t]
\vspace*{-5.7cm}
\centerline{\hspace*{4.cm}\psfig{file=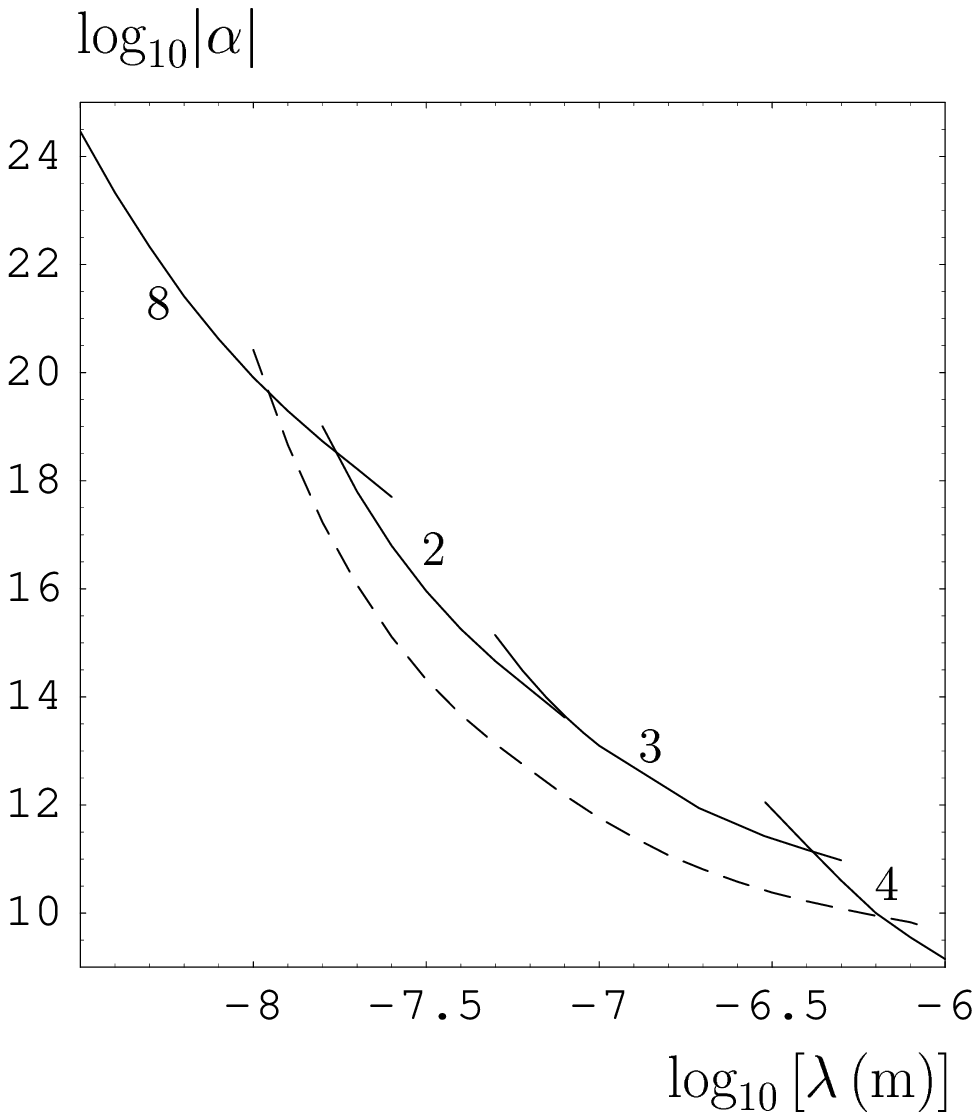,width=10cm}}
\vspace*{-3.2cm}
\caption{The prospective constraints on Yukawa-type interactions
which can be obtained from measurements of the gradient of normal
Casimir force  between an Au-coated microfabricated cylinder and
an Au-coated plate are shown by the dashed line. Lines 2, 3, 4,
and 8 are the same as in Figs.~1 and 4.
}
\end{figure*}
%%%%%%%%%%%%%%%

Equation (\ref{eq21}) and its generalizations were used\cite{42} to derive
the strongest constraints on Yukawa-type interactions that could be
obtained from measurements of the gradient of the Casimir force
between a microfabricated cylinder and a plate. The derived constraints
are shown by the dashed line in Fig.~7. The other lines show the
strongest constraints obtained from measurements of the lateral Casimir
force (line 8), from measurements of the Casimir pressure by means of
a micromachined oscillator (line 2), from the Casimir-less
experiment (line 3) and from the torsion-pendulum experiment of 1997
(line 4). As can be seen in Fig.~7, the proposed experiment promises to
strengthen the currently available constraints over a very wide
interaction range from $\lambda=12.5\,$nm to $\lambda=630\,$nm.
The strongest strengthening, up to 70 times, can be achieved at
$\lambda=18\,$nm. This makes the cylinder-plate geometry promising not only
for measurements of the Casimir force, but for obtaining new constraints
on non-Newtonian gravity as well.

\section{Conclusions}

As can be concluded from the foregoing, experiments on the Casimir force
lead to strongest constraints on Yukawa-type interactions at shorter
interaction range where the
gravitational experiments fail to provide competitive
constraints. One can also conclude that recent measurements of the
normal Casimir force between a sphere and a plate with trapezoidal
corrugations and of the thermal Casimir-Polder force qualitatively
confirm constraints obtained from earlier experiments. As to measurements
of the lateral Casimir force, it resulted in the strengthening of the
previously known high-confidence constraints up to a factor of 24
millions. So strong strengthening has no precedent in other
experiments.
To conclude, in near future further strengthening of constraints on
Yukawa interaction is expected from measuring the Casimir force in
configurations with corrugated boundaries and in cylindrical geometries.

\section*{Acknowledgments}

V.M.M.\ and G.L.K.\ were partially supported by the   NSF Grant
No.~PHY0970161 and by the DFG Grant BO\ 1112/20-1.
V.B.B.\ and C.R.\ were partially supported by CNPq.

%%%%%%%%%%%%%%%%%%%%%%%%%%%%%%%%%%%%%%%

\end{document}